\documentclass[letterpaper,twocolumn,10pt]{article}

\usepackage{usenix-2020-09}

\usepackage{graphicx}

\usepackage[cmex10]{amsmath}

\usepackage{amsthm}

\usepackage{makecell}

\usepackage{subcaption}

\usepackage{xcolor}

\newcommand{\parheader}[1]{\vspace{.3em}\noindent{}\textbf{{#1}}}
\newcommand{\eg}{e.g.,}
\newcommand{\ie}{i.e.,}
\newcommand{\etc}{etc.}

\newcommand{\wrt}{w.r.t.}
\newcommand{\ifandonlyif}{\textit{iff}}

\newcommand{\toolseqnature}{\textsc{Seqnature}}
\newcommand{\toolpingpong}{PingPong}

\newcommand{\smarttv}{smart TV}
\newcommand{\smarttvtitlecase}{Smart TV}
\newcommand{\appletv}{Apple TV}
\newcommand{\firetv}{Fire TV}
\newcommand{\roku}{Roku}

\newcommand{\iotdefinition}{Internet of Things}
\newcommand{\iot}{IoT}
\newcommand{\tcpdump}{tcpdump}

\newcommand{\fqdndefinition}{fully qualified domain name}

\newcommand{\fqdn}{FQDN}
\newcommand{\eslddefinition}{effective second-level domain}
\newcommand{\esld}{eSLD}
\newcommand{\http}{HTTP}
\newcommand{\dns}{DNS}
\newcommand{\tcp}{TCP}
\newcommand{\tls}{TLS}

\newcommand{\dnsoverhttps}{DoH}

\newcommand{\dnsovertls}{DoT}
\newcommand{\snidefinition}{Server Name Indication}
\newcommand{\sni}{SNI}

\newcommand{\natdefinition}{Network Address Translation}
\newcommand{\nat}{NAT}

\newcommand{\trafficsample}{traffic sample}
\newcommand{\trafficsamplesentencecase}{Traffic sample}
\newcommand{\trafficsamples}{traffic samples}
\newcommand{\tabtrafficsample}{tabulated traffic sample}
\newcommand{\tabtrafficsamples}{tabulated traffic samples}
\newcommand{\symboltrafficsamples}{\ensuremath{T}}
\newcommand{\symbolstreamprefixlength}{\ensuremath{P}}
\newcommand{\symbolmintrafficsamples}{\ensuremath{T_{\mathrm{min}}}}
\newcommand{\symbolminpktseqlen}{\ensuremath{n_{\mathrm{min}}}}
\newcommand{\fofapp}[1]{$F(#1)$}

\newcommand{\sdbfdefinition}{size and direction-based fingerprint}

\newcommand{\sdbf}{SDBF}
\newcommand{\esdbfdefinition}{endpoint, size, and direction-based fingerprint}

\newcommand{\esdbf}{ESDBF}
\newcommand{\ebfdefinition}{endpoint-based fingerprint}

\newcommand{\ebf}{EBF}
\newcommand{\fqdnbfdefinition}{\fqdn{}-based fingerprint}

\newcommand{\fqdnbf}{\fqdn{}BF}
\newcommand{\esldbfdefinition}{\esld{}-based fingerprint}

\newcommand{\esldbf}{\esld{}BF}
\newcommand{\dbfdefinition}{domain-based fingerprint} 
\newcommand{\dbf}{DBF}
\newcommand{\pbfdefinition}{packet-pair-based fingerprint}
\newcommand{\pbf}{PBF}

\newcommand{\plsdefinition}{packet-level signature} 
\newcommand{\pls}{PLS}

\newcommand{\datasetfingerprintv}{FingerprinTV}
\newcommand{\datasetpingpong}{PingPong}

\theoremstyle{definition}
\newtheorem{definition}{Definition}[section]

\begin{document}

\date{}

\title{\Large \bf \toolseqnature{}: Extracting Network Fingerprints from Packet Sequences}

\author{
{\rm Janus Varmarken}\thanks{This work was conducted when the author was a student at the University of California, Irvine. The views expressed in this paper are those of the authors, not Juniper Networks, Inc.}\\
Juniper Networks, Inc.
\and
{\rm Rahmadi Trimananda}\\
University of California, Irvine
\and
{\rm Athina Markopoulou}\\
University of California, Irvine
} 

\maketitle

\begin{abstract}
This paper proposes a general network fingerprinting framework, \toolseqnature{}, that uses packet sequences as its basic data unit and that makes it simple to implement \emph{any} fingerprinting technique that can be formulated as a problem of identifying packet exchanges that consistently occur when the fingerprinted {\em event} is triggered.
We demonstrate the versatility of \toolseqnature{} by using it to implement five different fingerprinting techniques, as special cases of the framework, which broadly fall into two categories: (i) fingerprinting techniques that consider features of each individual packet in a packet sequence, \eg{} size and direction; and (ii) fingerprinting techniques that only consider stream-wide features, specifically what Internet endpoints are contacted.
We illustrate how \toolseqnature{} facilitates comparisons of the relative performance of different fingerprinting techniques by applying the five fingerprinting techniques to datasets from the literature.
The results confirm findings in prior work, for example that endpoint information alone is insufficient to differentiate between individual events on \iotdefinition{} devices, but also show that \smarttv{} app fingerprints based exclusively on endpoint information are not as distinct as previously reported.
\end{abstract}

\section{Introduction}
\label{sec:seqnature-overview}
Network fingerprinting (\ie{} identifying applications, events, \etc{} based on features extracted from network traffic) has a long history and applications in the fields of network measurement, security and privacy, and across application domains.
For example, motivated by the security and privacy concerns arising from the proliferation of \iotdefinition{} (\iot{}) devices in the wild over the past decade, recent work has shown that events on \iot{} devices can be identified from the metadata of sequences of packets exchanged between the \iot{} device and the device vendor's server(s)~\cite{oconnor2019homesnitch, trimananda2020}.

In this paper, we build on this idea and propose a general network fingerprinting framework, \toolseqnature{}, that uses packet sequences as its basic data unit and that makes it simple to implement the extraction of any arbitrary fingerprint of interest, provided that the extraction can be formulated as a problem of identifying consistently occurring packet sequences.
\toolseqnature{} is platform-agnostic, \ie{} \toolseqnature{} can be used to identify fingerprints of events on any software/hardware platform, including, but not limited to, \iot{}.

A fingerprint for event $e$, as extracted by \toolseqnature{}, is the set of packet sequences that consistently occur when $e$ is triggered.
A \emph{packet sequence}, formally defined in Definition~\ref{def:seqnature-packet-sequence}, is an $n$-gram of packets in a \tcp{} stream.
Alternatively, a packet sequence can be thought of as analogous to a substring, but where packets assume the role of characters.

\begin{definition}
\label{def:seqnature-packet-sequence}
A \emph{packet sequence} of length $n$ is $n$ consecutive packets within a \tcp{} stream, after retransmissions and zero-payload packets have been dropped and the packets in the \tcp{} stream have been ordered by timestamp.
\end{definition}

\toolseqnature{}'s configuration parameters allow the user to define when packet sequences are considered identical, and to constrain the lengths of packet sequences considered for inclusion in the fingerprint.
\toolseqnature{} therefore enables the implementation of complex fingerprinting techniques that consider features of each individual packet in a packet sequence (\eg{} size and direction, as in \plsdefinition{}s~\cite{trimananda2020}), but can also be used to implement simpler fingerprinting techniques that only consider stream-wide features (\eg{} the set of Internet endpoints contacted, as in \dbfdefinition{}s~\cite{varmarken2022fingerprintv}).
By making it simple to implement (and combine) existing and new fingerprinting techniques, \toolseqnature{} can facilitate studies of the relative performance of different fingerprinting techniques.

\begin{figure*}[t!]
    \centering
    \includegraphics[width=0.7\linewidth]{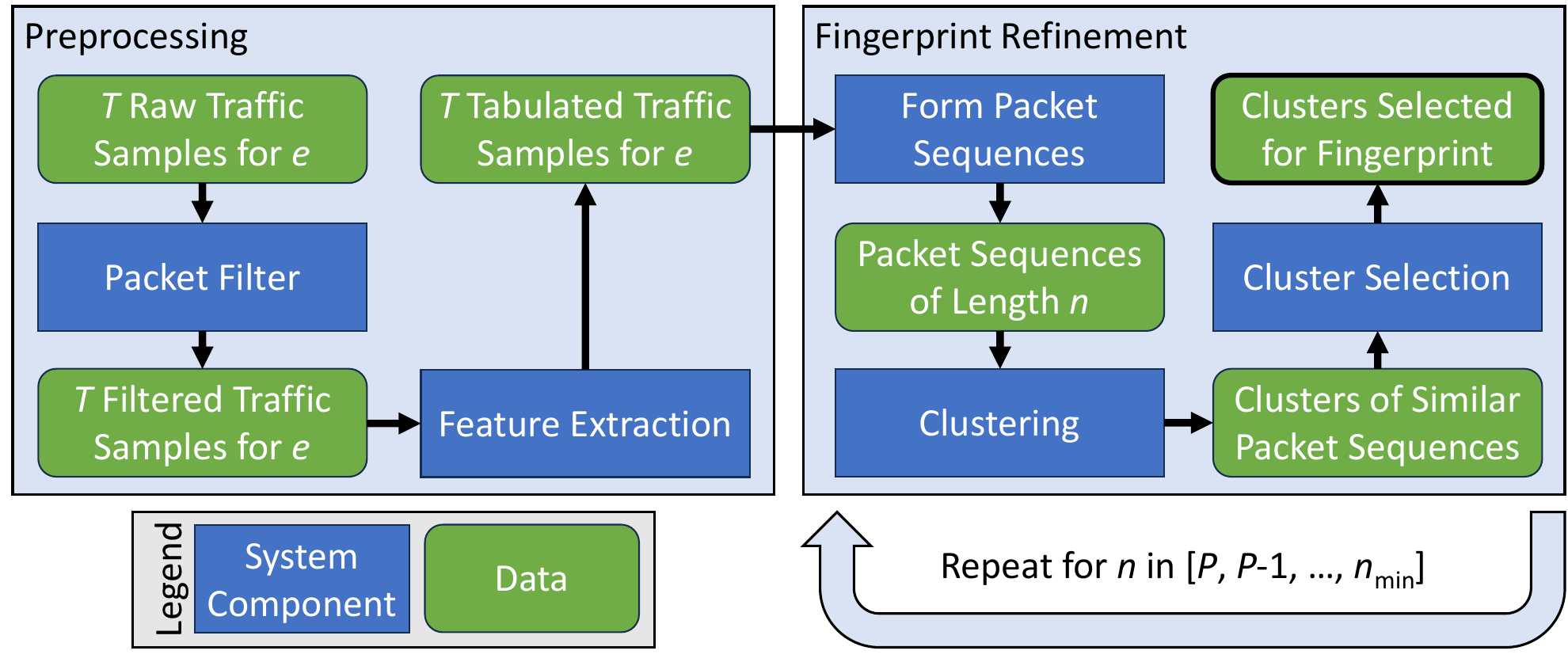}
    \caption{Overview of \toolseqnature{}. To fingerprint event $e$, \toolseqnature{} is provided with \symboltrafficsamples{} samples of the network traffic that occurred immediately after $e$ was triggered. \toolseqnature{} has two phases: a preprocessing phase (Section~\ref{sec:seqnature-fingerprinting-technique-preprocessing}) that extracts TCP stream information from the raw traffic samples, and an iterative fingerprint refinement phase (Section~\ref{sec:seqnature-fingerprinting-technique-refinement}) that identifies packet sequences that co-occur with $e$.}
    \label{fig:seqnature-system-overview}
\end{figure*}

\parheader{Fingerprinting Framework.}
An overview of \toolseqnature{} is provided in Figure~\ref{fig:seqnature-system-overview}.
To fingerprint event $e$ on computing device $d$, \toolseqnature{} is provided with \symboltrafficsamples{} samples of the network traffic that occurred on $d$ immediately after $e$ was triggered.
These raw packet captures are referred to as \emph{\trafficsamples{}}.
The \trafficsamples{} are first reduced to only the \tcp{} packets to/from $d$ that carry payload.
The filtered \trafficsamples{} are then converted to \emph{\tabtrafficsamples{}} by transforming each remaining packet into a vector of features, where the features are the server the packet was exchanged with, the packet's size, the packet's direction \etc{}

The core of \toolseqnature{} is an iterative procedure, referred to as \emph{fingerprint refinement}, which relies on clustering to identify packet sequences that consistently co-occur with $e$.
In order to identify consistently occurring packet sequences of any possible length $n$, fingerprint refinement iteratively considers increasingly shorter packet sequences.
Fingerprint refinement can be customized using a range of parameters.
This allows for flexibility in terms of when two packet sequences observed in different \trafficsamples{} are considered the same.
For example, the parameters can be used to specify the allowed variance in packet sizes.

When fingerprint refinement concludes, it outputs a set of clusters of packet sequences which constitutes the final fingerprint.
This fingerprint is also referred to as the \emph{seqnature} of $e$, a play on the words ``sequence'' and ``signature'', the latter of which is sometimes used as a synonym for fingerprint.
Throughout the remainder of this paper, we therefore use fingerprint and seqnature interchangeably when referring to the fingerprint of some event $e$.
Each cluster in the resulting seqnature represents a group of packet sequences that are considered identical, and which appear across a user-defined minimum number of different \trafficsamples{}.
The seqnature can be used to identify the occurrence of $e$ in unseen traffic by checking if, for every cluster in the seqnature, there exists at least one packet sequence in the candidate traffic that would fall in that cluster.

\parheader{Contributions.}
The primary contribution of this paper is \toolseqnature{}, a fingerprinting framework that makes it simple to implement, and evaluate the performance of, any fingerprinting technique that extracts fingerprints that are based on packet sequences.
To demonstrate the versatility of \toolseqnature{}, we use it to implement five well-known fingerprinting techniques, as special cases of the general framework. These extract five different types of fingerprints, which broadly fall in two categories: (i) fingerprints that consider the amount of data being exchanged in both directions, at the granularity of the individual packet; and (ii) fingerprints that consider only the set of contacted Internet endpoints. 
We plan to make the source code for \toolseqnature{} available.

As a secondary contribution, we provide a comparison of the performance of the aforementioned fingerprinting techniques in the context of \smarttv{} apps and events on simple \iot{} devices, such as smart plugs and smart light bulbs, by applying the five fingerprinting techniques to publicly available datasets from the literature~\cite{varmarken2022fingerprintv, trimananda2020}.
Our results confirm findings in prior work, for example that endpoint information alone is insufficient to differentiate between individual events on \iot{} devices, but also show that \smarttv{} app fingerprints based exclusively on endpoint information are not as distinct as previously reported. The comparison is primarily meant to illustrate how \toolseqnature{} can be used to formulate different fingerprinting techniques in a unifying way, and facilitate their comparison.

\parheader{Outline.}
The remainder of this paper is structured as follows.
Section~\ref{sec:seqnature-related-work} sumarizes related work.
Section~\ref{sec:seqnature-fingerprinting-framework} describes the design and implementation of the \toolseqnature{} framework.
Section~\ref{sec:seqnature-fingerprinting-techniques} defines the five fingerprinting techniques we consider and explains how they are implemented using \toolseqnature{}.
Section~\ref{sec:seqnature-datasets} introduces the datasets we use in our evaluation of the five fingerprinting techniques.
Section~\ref{sec:seqnature-fingerprinting-results} compares the performance of the five fingerprinting techniques.
Section~\ref{sec:seqnature-limitations-and-future-directions} discusses limitations and future directions.
Section~\ref{sec:seqnature-summary} concludes the paper.

\section{Related Work}
\label{sec:seqnature-related-work}

Network fingerprinting (\ie{} identifying applications, events,
\etc{} based on features extracted from network traffic) has been studied extensively for a long time in the network measurement literature, as it informs network management, including traffic engineering and prioritization, network provisioning and intrusion detection. Conversely, it may also introduce privacy risks as it allows for passive observation of users' network usage and inference of usage pattern, ultimately revealing information on how they use and interact with their device and Internet services.

Early fingerprinting techniques proposed by the research community include those that relied on inspection of packet payload to fingerprint network applications~\cite{moore2005, ma2006}. The community then extended this technique to
emerging devices, such as mobile~\cite{miskovic2015}, IoT~\cite{feng2018}, etc. Unfortunately, the proposed techniques and findings in this body of work quickly became irrelevant with the rise of encryption~\cite{razaghpanah2017, alrawi2019, huang2020}.

\parheader{Encrypted Traffic Fingerprinting.}
With the widespread use of encryption, the research community then turned their attention to exploring new techniques to fingerprint encrypted traffic. Some have demonstrated that the application layer protocol can be fingerprinted using the sizes and directions of just a few initial packets of an SSL session~\cite{bernaille2007}.
Others have fingerprinted HTTP User-Agent strings~\cite{husak2016} and websites~\cite{hintz2002, sun2002}.
Some have explored techniques to fingerprint websites protected by various privacy enhancing technologies~\cite{oh2019, bhat2019, sirinam2019, oh2021, smith2021, ma2021, hoang2021}, and desktop~\cite{anderson2020accurate} and mobile~\cite{taylor2016, taylor2018, ede2020} applications have also been fingerprinted.
Furthermore, there have been techniques that fingerprint user actions by analyzing the traffic of mobile applications~\cite{conti2016, saltaformaggio2016}, voice assistants~\cite{kennedy2019, wang2020, charyyev2020, hyland2021, ahmed2023spying}, IoT devices~\cite{copos2016, oconnor2019, ren2019, trimananda2020, acar2020, dong2023horuseye, hu2023behaviot}, and even video content~\cite{reed2017, schuster2017}.
Generally, much of the recent work on traffic fingerprinting has embraced the power of (and advances in) machine learning~\cite{holland2021new, li2022packet}, and deep learning~\cite{qu2023input, xie2023rosetta}, to mention only some representative examples.

\parheader{Our Work in Perspective.}
\toolseqnature{} builds upon prior work that has proposed fingerprinting techniques that use endpoint information and/or packet metadata (\eg{} packet sizes and directions) as features~\cite{varmarken2022fingerprintv, trimananda2020,oconnor2019homesnitch}.
In contrast to these prior works, each of which propose a single (or a few~\cite{varmarken2022fingerprintv}), specific fingerprinting technique and demonstrates its power in a certain software/hardware context, \toolseqnature{} is a framework that makes it simple to implement and compare many -- new as well as (combinations of) existing -- fingerprinting techniques across different software/hardware contexts.

\section{Fingerprinting Framework}
\label{sec:seqnature-fingerprinting-framework}
In this section, we describe the design and implementation of \toolseqnature{}, a framework for extracting a network fingerprint for any arbitrary event $e$ that triggers network activity on a computing device $d$.
\toolseqnature{} has two phases (see Figure~\ref{fig:seqnature-system-overview}): (i) a \emph{preprocessing} phase, described in Section~\ref{sec:seqnature-fingerprinting-technique-preprocessing}, that extracts all TCP streams from $e$'s \trafficsamples{} and transforms packets into feature vectors; and (ii) an iterative \emph{fingerprint refinement} phase, described in Section~\ref{sec:seqnature-fingerprinting-technique-refinement}, that, by clustering increasingly shorter packet sequences, identifies packet sequences that consistently appear when $e$ occurs.
The final fingerprint of $e$, also referred to as the seqnature of $e$, is the set of these packet sequences, accompanied by information about how they may vary, as described in Section~\ref{sec:seqnature-fingerprint-representations}.
Section~\ref{sec:seqnature-fingerprint-matching} describes a basic algorithm we implement to examine a \tabtrafficsample{} for the manifestation of a given seqnature.
We use this algorithm to examine network traffic for false positives, \ie{} when the seqnature of $e$ manifests itself in traffic that does not stem from $e$.

\subsection{Preprocessing}
\label{sec:seqnature-fingerprinting-technique-preprocessing}
The preprocessing phase of \toolseqnature{} is summarized in the ``Preprocessing'' box in Figure~\ref{fig:seqnature-system-overview}.
This phase consists of two steps: (i) a step that filters packets in the raw \trafficsamples{}; and (ii) a step that vectorizes packets in the raw \trafficsamples{} to create a format suitable for cluster analysis.
Below, we detail each step in the context of a single \trafficsample{}, but all \symboltrafficsamples{} \trafficsamples{} are subjected to the same procedure.
It is assumed that the \trafficsamples{} are collected on $d$'s gateway router and that $d$ is assigned an IP address from one of the private IP address spaces.

\parheader{Filtering.}
The first preprocessing step filters packets in the raw \trafficsample{} so as to only retain those packets that are relevant for the fingerprinting objective.
As the intended use of \toolseqnature{} is to identify fingerprints that are observable upstream from the home router's WAN port, the filter only retains IP datagrams to/from $d$, where either the source or the destination is a public IP address.
However, as most local networks are configured to use a local DNS resolver by default (the home gateway often assumes this role), DNS messages to/from $d$ are not subject to the public IP address requirement.   

For \tcp{} streams, retransmissions and zero-payload packets are dropped prior to cluster analysis to ensure that the identified fingerprint (if any) only comprises packet exchanges that carry data relevant to the application, but \emph{not} packet exchanges that are a result of the intricacies of TCP's operation under certain network conditions.
Ideally, this should make the resulting fingerprint network-agnostic, enabling a fingerprint derived from training data collected in one network to be used for event detection in another network.

The filtering logic may be changed to accommodate different fingerprinting objectives.
For example, to extract fingerprints that are only observable on the LAN, the filter should only retain IP datagrams to/from $d$ where the other endpoint's IP address is also in one of the private address spaces.

\begin{table*}[t!]
    \small
    \centering
    \setlength\tabcolsep{2.5pt} 
    \begin{tabular}{|r|r|r|c|r|r|c|}
        \hline
        \textbf{\makecell[c]{Event ID}} & \textbf{\makecell[c]{Sample ID}} & \textbf{\makecell[c]{Stream ID}} & \textbf{\makecell[c]{Domain}} & \textbf{\makecell[c]{Position in Stream}} & \textbf{\makecell[c]{Size}} & \textbf{\makecell[c]{Direction}} \\
        \scriptsize{\makecell{Event $e$ packet \\ was observed for}} & \scriptsize{\makecell{\trafficsamplesentencecase{} \\ packet appeared in}} & \scriptsize{\makecell{TCP stream packet \\ pertains to}} & \scriptsize{\makecell{Domain name of the server}} & \scriptsize{\makecell{Packet's position \\ within its \tcp{} stream}} & \scriptsize{\makecell{Packet's size \\ (bytes)}} & \scriptsize{\makecell{Packet's \\ direction \wrt{} $d$}} \\
        \hline
        28 & 1 & 2 & scribe.logs.roku.com & 1 & 583 & upstream \\
        28 & 1 & 2 & scribe.logs.roku.com & 2 & 1514 & downstream \\
        28 & 1 & 2 & scribe.logs.roku.com & ... & ... & ... \\
        28 & 1 & 2 & scribe.logs.roku.com & 9 & 97 & downstream \\
        28 & 1 & ... & ... & ... & ... & ...\\
        28 & 1 & 6 & tuner.pandora.com & 1 & 293 & upstream \\
        28 & 1 & 6 & tuner.pandora.com & 2 & 188 & upstream \\
        28 & 1 & 6 & tuner.pandora.com & 3 & 362 & downstream \\
        28 & 1 & ... & ... & ... & ... & ... \\
        \hline
    \end{tabular}
    \caption{Snippet of a \tabtrafficsample{} for the Roku app with ID=28 (``Pandora'') in the \datasetfingerprintv{} dataset~\cite{varmarken2022fingerprintv}. Each row represents a single packet.}
    \label{tab:seqnature-tabulated-traffic-sample-example}
\end{table*}

\parheader{Feature Extraction.}
The second preprocessing step selects all TCP streams from the filtered \trafficsample{} and transforms each packet into a feature vector.
An example output and a description of each feature is provided in Table~\ref{tab:seqnature-tabulated-traffic-sample-example}.
We refer to this format as a \emph{\tabtrafficsample{}} as it lends itself to presentation in table form and is suitable as input for cluster analysis (see Section~\ref{sec:seqnature-fingerprinting-technique-refinement}).
The extracted features (i) tie the packet to the TCP stream it appeared in (the combination of features Event ID, Sample ID, and Stream ID); (ii) label the packet with the domain name of the server it was exchanged with (Domain); (iii) order the packet relative to other packets in the same stream (Position in Stream); (iv) log the packet's size (Size); and (v) log the packet's direction relative to $d$~(Direction).

The value for the Domain feature is extracted from the TLS \snidefinition{} (\sni{}), if the TCP stream wraps a TLS stream that includes a Client Hello message where the \sni{} extension is present.
Otherwise, the server's IP address is matched against IP addresses returned in preceding DNS responses, and Domain is set to the name queried in the most recent DNS response in which there is an IP address match.
If the domain name also cannot be determined from DNS, the server's IP address is used for Domain as a last resort.

The value for the Position in Stream feature is the packet's position relative to the other packets in the stream, after retransmissions and packets without TCP payload have been discarded.
The position is derived from the packet's timestamp, rather than its sequence number, to form a bidirectional ordering as opposed to two direction-specific orderings.
Timestamp-based ordering should also make it easier to add support for UDP later on.

\subsection{Fingerprint Refinement}
\label{sec:seqnature-fingerprinting-technique-refinement}
The fingerprint refinement phase of \toolseqnature{} is summarized in the ``Fingerprint Refinement'' box in Figure~\ref{fig:seqnature-system-overview}.
This phase selects packet sequences for inclusion in the seqnature by identifying identical (or similar) packet sequences that consistently (or often) occur when $e$ is triggered.

The input to fingerprint refinement is the \symboltrafficsamples{} tabulated \trafficsamples{} resulting from the preprocessing phase described in Section~\ref{sec:seqnature-fingerprinting-technique-preprocessing}.
In summary, fingerprint refinement (i) forms all possible packet sequences of length $n$ from the TCP streams in the \symboltrafficsamples{} tabulated \trafficsamples{}; (ii) clusters said packet sequences to identify identical (or similar) ones; and (iii) post-processes the resulting clusters to select clusters for inclusion in the fingerprint based on criteria such as how many of the $T$ \trafficsamples{} the packet sequences in a cluster stem from.

Fingerprint refinement is an iterative process that considers increasingly shorter packet sequences, \ie{} the aforementioned steps are repeated for all $n \in [\symbolstreamprefixlength{}, \symbolstreamprefixlength{}-1, ..., \symbolminpktseqlen{}]$, where $\symbolstreamprefixlength{} \ge{} \symbolminpktseqlen{} \ge{} 1$.
The user controls \symbolstreamprefixlength{} and \symbolminpktseqlen{} using \toolseqnature{}'s command-line interface.
\symbolstreamprefixlength{} specifies how many packets of each \tcp{} stream should be considered during fingerprint refinement, and \symbolminpktseqlen{} specifies the minimum length of any packet sequence considered for inclusion in the seqnature.
Below, we describe each of the three fingerprint refinement steps in detail.

\begin{figure}[t!]
    \centering
    \includegraphics[width=\linewidth]{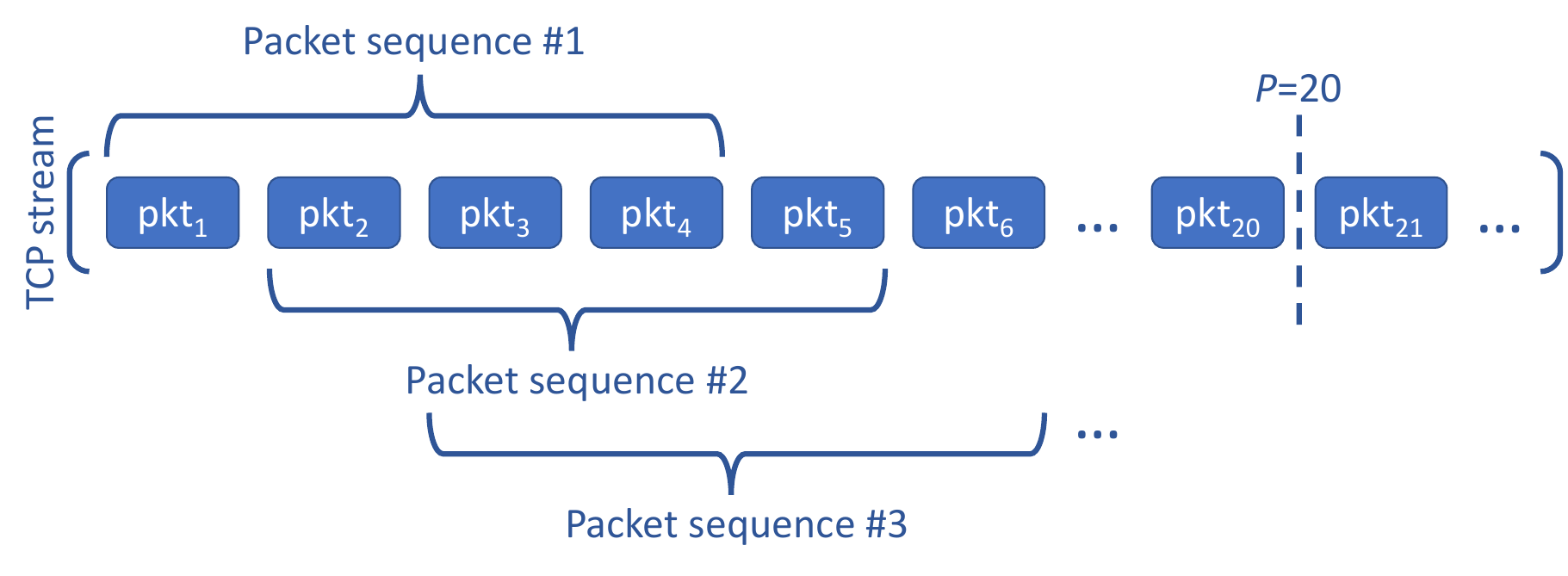}
    \caption{Example of how packet sequences of length $n=4$ are formed from the first $\symbolstreamprefixlength{}=20$ packets of a \tcp{} stream.}
    \label{fig:seqnature-forming-packet-sequences}
\end{figure}

\begin{figure*}[t!]
    \centering
    \includegraphics[width=0.7\linewidth]{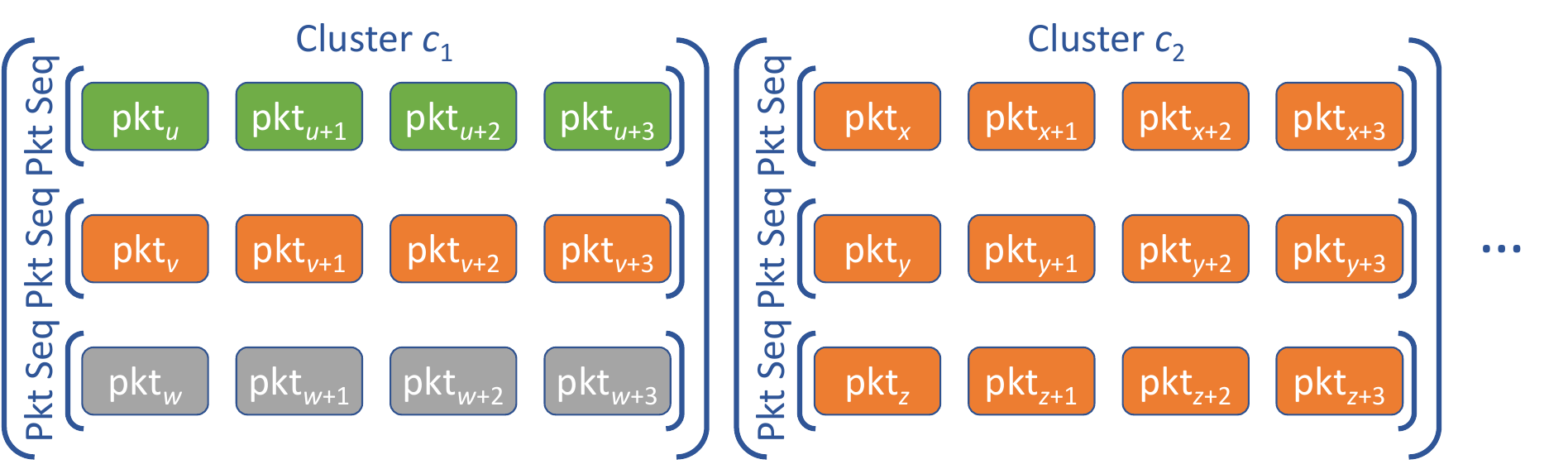}
    \caption{Example clustering of packet sequences of length $n=4$ extracted from $\symboltrafficsamples{}=3$ different \trafficsamples{}. The color of packets in a packet sequence denotes what \trafficsample{} the packet sequence stems from. For example, all packet sequences with orange packets stem from the same \trafficsample{}. The number of packet sequences in a cluster may vary across clusters (and can be greater than $\symboltrafficsamples{}$), but all packet sequences across all clusters will always be of the same length $n$, as $n$ only changes between each fingerprint refinement iteration.}
    \label{fig:seqnature-clustering-example}
\end{figure*}

\parheader{Forming Packet Sequences.}
The first step of each fingerprint refinement iteration forms all possible packet sequences of length $n$ from the first \symbolstreamprefixlength{} packets of each \tcp{} stream in the \symboltrafficsamples{} tabulated \trafficsamples{}.
As a packet sequence is essentially an $n$-gram of packets in a \tcp{} stream (see Definition~\ref{def:seqnature-packet-sequence}), \toolseqnature{} forms all possible packet sequences of length $n$ by sliding a window of size $n$ across the first \symbolstreamprefixlength{} packets of each \tcp{} stream.
An example of how packet sequences of length $n=4$ are formed from the first $\symbolstreamprefixlength{}=20$ packets of a \tcp{} stream is provided in Figure~\ref{fig:seqnature-forming-packet-sequences}.

The parameter \symbolstreamprefixlength{} ensures that fingerprint refinement remains computationally tractable for \tcp{} streams with many packets, \eg{} \tcp{} streams carrying large data volumes.
Throughout this paper, we set $\symbolstreamprefixlength{}=20$ unless otherwise stated.
This choice is grounded in prior work which has demonstrated that a small prefix of packets of each stream suffices for classifying \iot{} traffic, mobile app traffic, and \dnsovertls{}/\dnsoverhttps{} traffic~\cite{rezaei2019overview, rezaei2020mobileapp, siby2020encrypteddns, lopezmartin2017cnnrnn, anderson2017malware}.
Using a small value for \symbolstreamprefixlength{} also makes online fingerprint detection more feasible in comparison to approaches that need access to all packets in the stream, \eg{} those that rely on statistical measures of packet-level features such as packet size.

\parheader{Clustering.}
The second step of each fingerprint refinement iteration clusters the packet sequences formed in the preceding step to group identical (or similar) packet sequences together based on a user-defined notion of packet sequence similarity.
The resulting clusters become candidates for inclusion in the fingerprint as they \emph{may} represent packet exchanges that co-occur with $e$.
An example clustering of packet sequences of length $n=4$ extracted from $\symboltrafficsamples{}=3$ different \trafficsamples{} is provided in Figure~\ref{fig:seqnature-clustering-example}.

Clustering in \toolseqnature{} is fully customizable: support for any clustering algorithm available in scikit-learn~\cite{sklearnmain} (or any Python module that adopts its API~\cite{sklearnapi}, \eg{} hdbscan~\cite{mcinnes2017hdbscan}) can be added with two lines of code; parameters for configuring the algorithm of choice can be specified using \toolseqnature{}'s command-line interface; and any custom distance metric can be introduced by defining a function in a dedicated module.
This configurability leaves the user in full control of what packet sequences end up in the same cluster, which in turn allows for the implementation of many different fingerprinting techniques, as demonstrated in Section~\ref{sec:seqnature-fingerprinting-techniques}.

\parheader{Cluster Selection.}
The third fingerprint refinement step post-processes the clusters formed in the previous step to determine which, if any, should be included in the fingerprint for $e$.
A cluster of identical (or similar) packet sequences is not necessarily useful as one part of the fingerprint for $e$ as there is no guarantee that all \symboltrafficsamples{} invocations of $e$ are represented in the cluster.
For example, if the same packet sequence occurs \symboltrafficsamples{} times in a single \trafficsample{}, but not in any of the remaining $\symboltrafficsamples{}-1$ \trafficsamples{}, a cluster of \symboltrafficsamples{} packet sequences will be formed, but such a cluster is obviously not useful for fingerprinting purposes (cluster $c_2$ in Figure~\ref{fig:seqnature-clustering-example} illustrates this scenario).
For this reason, \toolseqnature{} defines a parameter, \symbolmintrafficsamples{}, that is used to specify how many different \trafficsamples{} packet sequences in a cluster must stem from in order for the cluster to be considered valid for inclusion in the fingerprint for $e$.
Throughout this paper, we set $\symbolmintrafficsamples{}=\symboltrafficsamples{}$, \ie{} the strictest case where a packet sequence must always co-occur with $e$ for the packet sequence to become part of the seqnature of $e$.

\begin{figure*}[t!]
    \centering
    \includegraphics[width=0.6\linewidth]{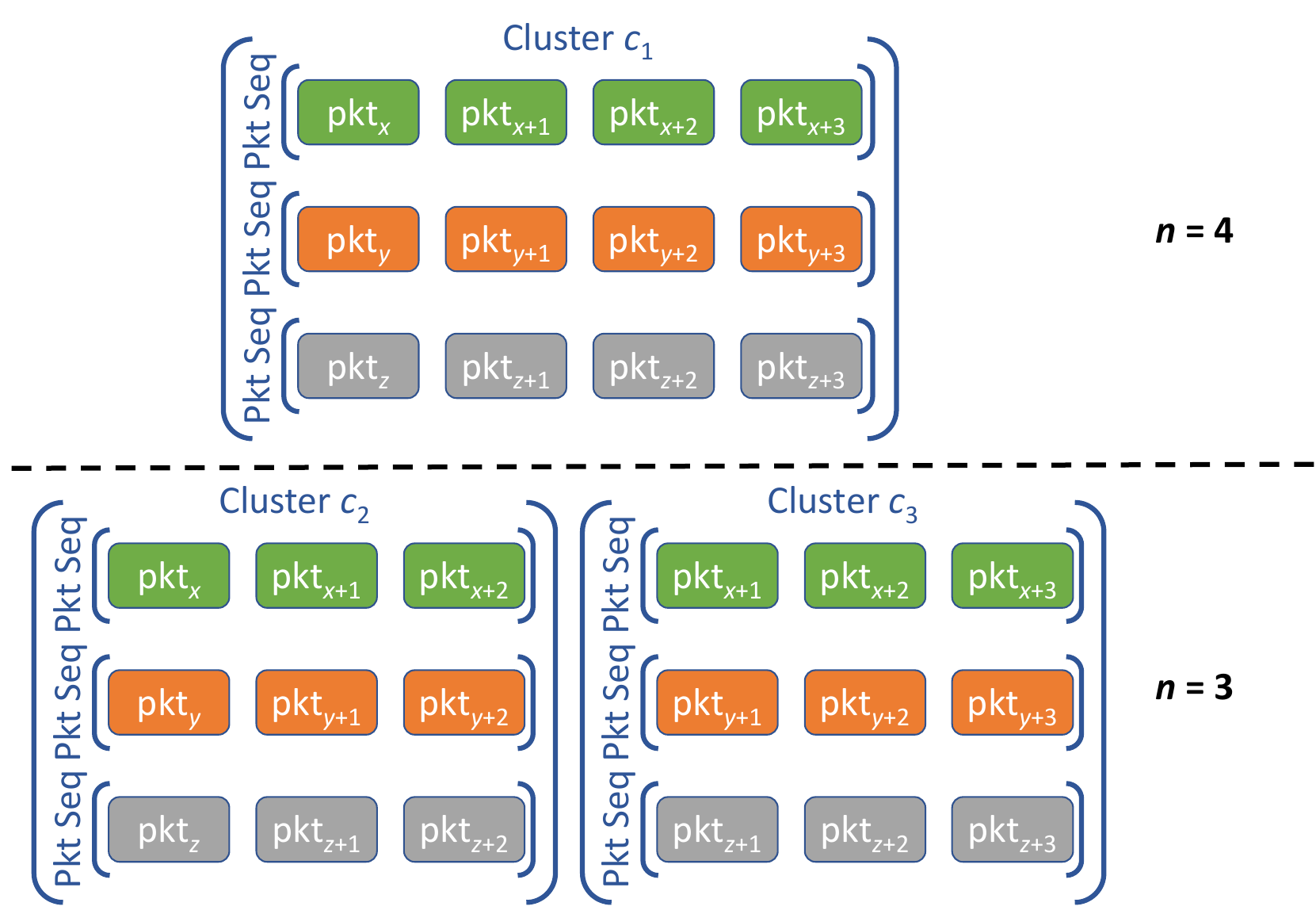}
    \caption{Example clusterings for two successive fingerprint refinement iterations ($n=4$ and $n=3$). When $n$ decreases, at least one cluster containing shorter versions of packet sequences that have already been included in the seqnature will be formed: in this example, clusters $c_2$ and $c_3$ both (exclusively) consist of shorter versions of the packet sequences in cluster $c_1$.}
    \label{fig:seqnature-cluster-selection-duplicates}
\end{figure*}

Finally, each valid cluster is compared against clusters already included in the fingerprint for $e$ to prevent duplicates.
Specifically, if a cluster of packet sequences of length $n_j$ has been identified as valid and included in the fingerprint, all subsequent fingerprint refinement iterations operating on packet sequences of length $n_i$, where $n_i < n_j$, will give rise to at least one cluster of packet sequences of length $n_i$ that are shorter versions of the packet sequences of length $n_j$ (see Figure~\ref{fig:seqnature-cluster-selection-duplicates} for an example).
Naturally, such clusters should not be included in the fingerprint.
Therefore, before including a valid cluster $c$ in the fingerprint, \toolseqnature{} first compares all packet sequences in $c$ against the packet sequences in the clusters that are already in the fingerprint.
If any packet sequence $s$ in $c$ is already accounted for in the fingerprint, in the sense that a longer version of $s$ exists in one of the clusters in the fingerprint, $c$ is discarded.
To make packet sequence lookup more efficient, each cluster in the fingerprint is stored as a generalized suffix tree of the packet sequences it contains~\cite{mccreight1976, ukkonen1995, perathonersuffixtreegithub}.

\begin{figure*}[t!]
    \centering
    \begin{subfigure}{0.54\linewidth}
        \centering
        \includegraphics[width=\linewidth]{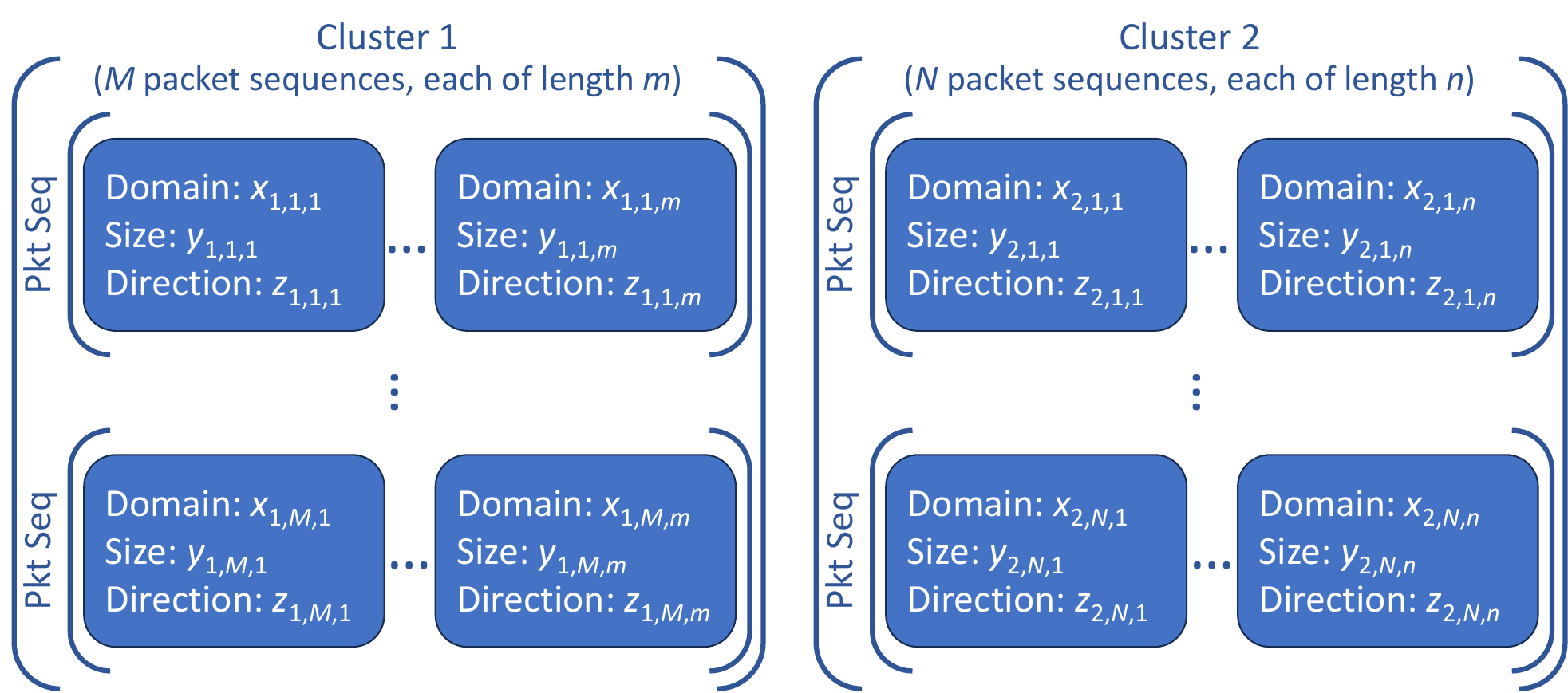}
        \caption{Complete form.}
        \label{fig:seqnature-seqnature-representations-complete}
    \end{subfigure}
    \hfill
    \begin{subfigure}{0.43\linewidth}
        \centering
        \includegraphics[width=\linewidth]{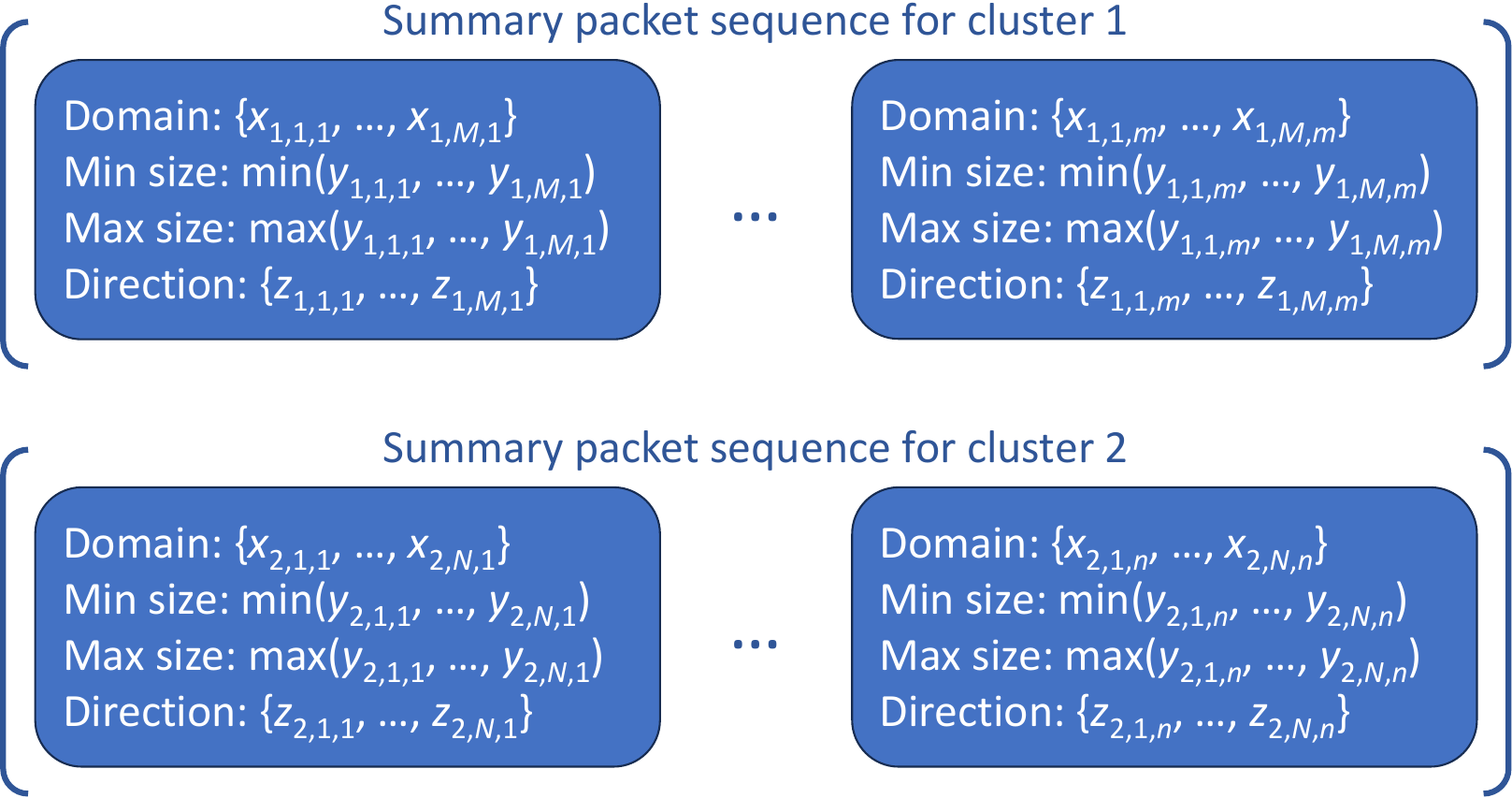}
        \caption{Summary form.}
        \label{fig:seqnature-seqnature-representations-summary}
    \end{subfigure}
    \caption{Example of a seqnature that comprises two clusters, represented in complete form and in summary form. Subscript notation: first value is the cluster number, second value is the packet sequence number (within the cluster), and third value is the packet index (within the packet sequence).}
    \label{fig:seqnature-seqnature-representations}
\end{figure*}

\subsection{Representations of the Seqnature}
\label{sec:seqnature-fingerprint-representations}
The final fingerprint for $e$, also referred to as the seqnature of $e$, is the set of clusters selected for inclusion in the fingerprint when fingerprint refinement concludes, \ie{} when $n < \symbolminpktseqlen{}$.
The seqnature can be used to identify the occurrence of $e$ in unseen traffic by checking if, for every cluster in the seqnature, there exists at least one packet sequence in the candidate traffic that would fall in that cluster.
To simplify the implementation of such an algorithm, and to make visual inspection of seqnatures easier, we condense the resulting seqnature by replacing each cluster of packet sequences with a single summary packet sequence.
We describe this summarization below and provide an example of the same seqnature before and after summarization in Figure~\ref{fig:seqnature-seqnature-representations}.

Since $n$ only changes between each fingerprint refinement iteration, each cluster in the final seqnature will only contain packet sequences of the same length.
Therefore, we can summarize each cluster $c$ as a single packet sequence $p$ in which the packet at index $i$, $p[i]$, encodes, for each feature $f$, the range of values $f$ assumes for packet $i$ across all packet sequences in $c$.
For some features, this can simplify the representation significantly.
For example, for packet size, we only store the minimum and the maximum size observed across all packets at index $i$.
For other features, such as the server's domain name and the packet's direction, we must store the set of values observed across all packets at index $i$ in order to accommodate user-defined distance metrics that do not require these features to assume identical values, \eg{} a distance metric that only requires the \eslddefinition{}s (\esld{}s), but not the \fqdndefinition{}s (\fqdn{}s), to match.

\subsection{Fingerprint Matching}
\label{sec:seqnature-fingerprint-matching}
\toolseqnature{} includes functionality for examining a \tabtrafficsample{} for the manifestation of a seqnature represented in summary form (see Section~\ref{sec:seqnature-fingerprint-representations}).
This functionality is used in Section~\ref{sec:seqnature-fingerprinting-results-distinctiveness}, where we evaluate the distinctiveness of different types of fingerprints. 
The matching algorithm declares a match of seqnature $s$ in \tabtrafficsample{} $t$ if every summary packet sequence in $s$ has at least one match in $t$.
We briefly describe the algorithm below.

To limit memory usage, the algorithm reads and processes $t$ packet-by-packet, as opposed to reading all packets in $t$ into memory at once.
For each \tcp{} stream in $t$, the algorithm maintains a buffer of the $n_{\mathrm{max}}$ most recent packets, where $n_{\mathrm{max}}$ is the length of the longest summary packet sequence in $s$.
Whenever a new packet is read from $t$ and added to its respective \tcp{} stream's buffer $b$, every summary packet sequence $p$ in $s$, for which no match has been found in $t$, is compared to the $n$ most recent packets in $b$, where $n$ is the number of packets in $p$ and $n \le n_{\mathrm{max}}$.
In order to facilitate detection of any arbitrary, user-defined type of seqnature, \toolseqnature{} allows the user the option to define their own custom logic for when the $n$ most recent packets in $b$ is considered a match of $p$.
The algorithm terminates when either (i) a match has been found in $t$ for all packet sequences in $s$, in which case there is a match of $s$ in $t$; or (ii) all packets in $t$ have been processed, in which case there is no match of $s$ in $t$.

\section{Fingerprinting Techniques}
\label{sec:seqnature-fingerprinting-techniques}
In this section, we demonstrate the versatility of \toolseqnature{} by using it to implement five different fingerprinting techniques. 
The fingerprinting techniques identify fingerprints that broadly fall into two categories: (i) fingerprints based on consistently occurring data exchanges; and (ii) fingerprints based on what Internet endpoints are consistently contacted.
Taken together, these fingerprinting techniques illustrate how \toolseqnature{}'s configurability -- in particular its flexible notion of packet sequence similarity -- enables the user to easily implement diverse fingerprinting techniques.
This section describes the five fingerprinting techniques and how we configure \toolseqnature{} to realize them.
An evaluation of their performance is deferred to Section~\ref{sec:seqnature-fingerprinting-results}.

\subsection{Fingerprints Based on Data Exchanges}
The first category of fingerprints are based on the hypothesis that if the same amount of data is consistently exchanged (with the same endpoints) whenever $e$ occurs, then these data exchanges can be used to identify $e$, if they are distinct from other traffic.
Fingerprints that fall in this category range from simple fingerprints that only consider the total number of bytes exchanged over each \tcp{} stream, to more complex fingerprints that also consider directionality and order of the data exchanges.
We focus on the more complex fingerprints, but the simpler fingerprints can also be implemented easily using \toolseqnature{}.

\subsubsection{Size and Direction}
The first fingerprinting technique we implement extracts fingerprints that consist of packet sequences where the directions and sizes of packets at corresponding indices stay consistent across \trafficsamples{}.
We refer to such a fingerprint as a \emph{\sdbfdefinition{}} (\sdbf{}) and formalize it in Definition~\ref{def:seqnature-sdbf}.
\sdbf{}s exemplify how \toolseqnature{} can be used to easily implement (approximations of) existing fingerprinting techniques as they are identical to packet-level signatures, except for inter-sequence ordering~\cite{trimananda2020}.

\begin{definition}
\label{def:seqnature-sdbf}
The \emph{\sdbfdefinition{}} (\sdbf{}) of event $e$ is the set $S$ of packet sequences s.t. for every packet sequence $p$ in $S$, $p$ appears at least once in at least \symbolmintrafficsamples{} of the \symboltrafficsamples{} \trafficsamples{} for $e$.
Two packet sequences $p_1$ and $p_2$ are considered identical \ifandonlyif{} (i) $|p_1|=|p_2|$, \ie{} $p_1$ and $p_2$ are the same length; (ii) the directions of packets at corresponding indices in $p_1$ and $p_2$ are identical; and (iii) $\sum_{i=1}^{n}{\left| \left|p_1[i]\right|-\left|p_2[i]\right| \right|} \le h$, \ie{} the sum of absolute differences in sizes of packets at corresponding indices in $p_1$ and $p_2$ is below a user-defined threshold $h$.
\end{definition}

To enable extraction of \sdbf{}s using \toolseqnature{}, we define a distance metric that considers the distance between packet sequences $p_1$ and $p_2$ to be \emph{maximal} if any pair of packets at corresponding indices in $p_1$ and $p_2$ go in opposite directions.
Otherwise, the distance is $\sum_{i=1}^{n}{\left| \left|p_1[i]\right|-\left|p_2[i]\right| \right|}$, \ie{} the sum of absolute differences in sizes of packets at corresponding indices in $p_1$ and $p_2$.

In Section~\ref{sec:seqnature-fingerprinting-results}, we extract, and evaluate the performance of, the strictest possible case of \sdbf{}s, \ie{} when $\symbolmintrafficsamples{}=\symboltrafficsamples{}$ and $h=0$ (see Definition~\ref{def:seqnature-esdbf}).
In other words, a cluster must contain packet sequences from all $T$ \trafficsamples{} to be considered for inclusion in the \sdbf{}, and packet sequences must be identical to end up in the same cluster.
This strict case of \sdbf{}s is achieved by configuring \toolseqnature{} to use the DBSCAN clustering algorithm~\cite{ester1996dbscan} with parameters $\epsilon=0$ and $\mathrm{\texttt{MinPts}}=\symboltrafficsamples{}=10$.

\subsubsection{Endpoint, Size, and Direction}
The second fingerprinting technique we implement extends \sdbf{}s with endpoint information.
A fingerprint of this type, referred to as an \emph{\esdbfdefinition{}} (\esdbf{}) and formalized in Definition~\ref{def:seqnature-esdbf}, consists of those packet sequences where (i)~the endpoint and (ii)~the directions and sizes of packets at corresponding indices stay consistent across \trafficsamples{}.
\esdbf{}s illustrate how \toolseqnature{} makes it easy to experiment with the effects of combining fingerprinting techniques, as \esdbf{}s are similar to the combination of \dbfdefinition{}s and \pbfdefinition{}s discussed in~\cite{varmarken2022fingerprintv}.

\begin{definition}
\label{def:seqnature-esdbf}
The \emph{\esdbfdefinition{}} (\esdbf{}) of event $e$ is the set $S$ of packet sequences s.t. for every packet sequence $p$ in $S$, $p$ appears at least once in at least \symbolmintrafficsamples{} of the \symboltrafficsamples{} \trafficsamples{} for $e$.
Two packet sequences $p_1$ and $p_2$ are considered identical \ifandonlyif{} (i) $p_1$ and $p_2$ are exchanged with the same endpoint, where the endpoint is identified by its \fqdn{}, when available, or otherwise by its IP address; and (ii) all packet sequence equality requirements listed in Definition~\ref{def:seqnature-sdbf} are satisfied.
\end{definition}

Extraction of \esdbf{}s using \toolseqnature{} is made possible by extending the distance metric used for \sdbf{}s to also consider the distance between packet sequences $p_1$ and $p_2$ to be \emph{maximal} if $p_1$ and $p_2$ are exchanged with different endpoints.
Endpoints are considered different if they are identified by different \fqdn{}s, different IP addresses, or if one endpoint is identified by an \fqdn{} while the other is identified by an IP address.
To stay consistent with \sdbf{}s, in Section~\ref{sec:seqnature-fingerprinting-results}, we extract, and evaluate the performance of, the strictest possible case of \esdbf{}s, \ie{} when $\symbolmintrafficsamples{}=\symboltrafficsamples{}$ and $h=0$.
This is done by configuring \toolseqnature{} to perform clustering in the same way as for \sdbf{}s.

\subsection{Fingerprints Based on Endpoints}
In this section, we demonstrate how \toolseqnature{} can also be used to implement fingerprinting techniques that focus on stream-wide features.
We implement three such fingerprinting techniques and formalize the fingerprints they extract in Definitions~\ref{def:seqnature-ebf}, \ref{def:seqnature-fqdnbf}, and \ref{def:seqnature-esldbf}, respectively.
A fingerprint extracted using either of these three fingerprinting techniques is the set of endpoints that are consistently contacted whenever event $e$ occurs.
The fingerprints only differ in terms of how an endpoint is identified.

\begin{definition}
\label{def:seqnature-ebf}
The \emph{\ebfdefinition{}} (\ebf{}) of event $e$ is the set $S$ of endpoints s.t. for every endpoint $d$ in $S$, $d$ is contacted in at least \symbolmintrafficsamples{} of the \symboltrafficsamples{} \trafficsamples{} for $e$.
An endpoint is identified by its \fqdn{}, or, when no domain name information is available, by its IP address. 
\end{definition}

\begin{definition}
\label{def:seqnature-fqdnbf}
The \emph{\fqdnbfdefinition{}} (\fqdnbf{}) of event $e$ is the set $S$ of endpoints s.t. for every endpoint $d$ in $S$, $d$ is contacted in at least \symbolmintrafficsamples{} of the \symboltrafficsamples{} \trafficsamples{} for $e$.
An endpoint is identified by its \fqdn{}. 
\end{definition}

\begin{definition}
\label{def:seqnature-esldbf}
The \emph{\esldbfdefinition{}} (\esldbf{}) of event $e$ is the set $S$ of endpoints s.t. for every endpoint $d$ in $S$, $d$ is contacted in at least \symbolmintrafficsamples{} of the \symboltrafficsamples{} \trafficsamples{} for $e$.
An endpoint is identified by its \esld{}. 
\end{definition}

To enable extraction of \ebf{}s, \fqdnbf{}s, and \esldbf{}s using \toolseqnature{}, we define three similar distance metrics, where the distance between packet sequences $p_1$ and $p_2$ is 0 if $p_1$ and $p_2$ are exchanged with the same endpoint; otherwise, the distance is maximal.
The three distance metrics only differ in terms of when two endpoints are considered the same: endpoints $d_1$ and $d_2$ are the same if their identifiers, obtained as defined in Definitions~\ref{def:seqnature-ebf}, \ref{def:seqnature-fqdnbf}, and \ref{def:seqnature-esldbf}, are identical.
For example, if $p_1$ is exchanged with \url{x.example.com} and $p_2$ is exchanged with \url{y.example.com}, the distance metrics used for \ebf{}s and \fqdnbf{}s return maximal distance because the \fqdn{}s differ, but the distance metric used for \esldbf{}s returns 0 distance because the \fqdn{}s share the same \esld{}.

As the endpoint is the same for all packets in a stream, there is no need to analyze more than one packet per stream during fingerprint refinement.
Therefore, we use $\symbolstreamprefixlength{} = \symbolminpktseqlen{} = 1$ when extracting \ebf{}s, \fqdnbf{}s, and \esldbf{}s.
We configure \toolseqnature{} to use the DBSCAN algorithm with parameters $\epsilon=0$ and $\mathrm{\texttt{MinPts}}=\symboltrafficsamples{}=10$, which ensures that packet sequences will only end up in the same cluster if they are exchanged with the same endpoint.

In Section~\ref{sec:seqnature-fingerprinting-results}, we extract, and evaluate the performance of, \ebf{}s, \fqdnbf{}s, and \esldbf{}s when $\symbolmintrafficsamples{}=\symboltrafficsamples{}$.
In other words, we extract \ebf{}s, \fqdnbf{}s, and \esldbf{}s that only consist of the endpoints that are contacted in all \trafficsamples{} for the event $e$.
This is the most conservative choice as (i) it produces fingerprints that can be trusted to always be present when the event occurs; and (ii) the potential for false positives is greater as the set of endpoints that make up a fingerprint when $\symbolmintrafficsamples{}=\symboltrafficsamples{}$ will be a subset of the set of endpoints that make up the corresponding fingerprint when $\symbolmintrafficsamples{}<\symboltrafficsamples{}$ (intuitively, and as shown in~\cite{varmarken2022fingerprintv}, the more endpoints a fingerprint comprises, the more distinct it is from other traffic).

\section{Datasets}
\label{sec:seqnature-datasets}
To illustrate how \toolseqnature{} facilitates comparisons of different fingerprinting techniques, we use the fingerprinting techniques described in Section~\ref{sec:seqnature-fingerprinting-techniques} to extract fingerprints from two datasets.\footnote{The part of the PacketPrint~\cite{li2022packet} dataset that is captured from mobile apps using \tcpdump{} and used as ground truth meets \toolseqnature{}'s requirements (multiple \trafficsample{}s per app, and TCP/IP visibility), but unfortunately the authors could not release the raw PCAP files upon our request.}
Taken together, these datasets span different software and hardware categories: \smarttv{} apps and events on simple \iot{} devices, such as smart plugs and smart light bulbs.
This section introduces the datasets and describes how we preprocess them to make them compatible with \toolseqnature{}'s expected input format.
Section~\ref{sec:seqnature-fingerprinting-results} provides the results.


\subsection{\datasetfingerprintv{}: \smarttvtitlecase{} Apps}
\label{sec:seqnature-datasets-fingerprintv}
The first dataset we consider is the \datasetfingerprintv{} dataset, which is a recent dataset with network traffic from \smarttv{} apps~\cite{varmarken2022fingerprintv}.
In summary, the dataset consists of 10 samples of the network traffic that each of the top-1000 apps on \appletv{}, \firetv{}, and \roku{} generate at launch time, for a total of 30K \trafficsample{}s (10K per \smarttv{} platform).
Aside from the preprocessing performed as part of the fingerprinting procedure (see Section~\ref{sec:seqnature-fingerprinting-technique-preprocessing}), no preprocessing is necessary as this dataset is already in the format expected by \toolseqnature{}.



\subsection{\datasetpingpong{}: Events on \iot{} Devices}
\label{sec:seqnature-datasets-pingpong}
The second dataset we consider consists of network traffic that simple \iot{} devices, such as smart plugs and smart light bulbs, generate when an event is triggered, \eg{} when the user toggles a smart plug ON using the companion app of the smart plug.
We refer to this dataset as the \datasetpingpong{} dataset as it was published alongside~\cite{trimananda2020}, which proposed a fingerprinting technique of the same name.

The \datasetpingpong{} dataset is made available as one large packet capture (PCAP file) for each combination of \iot{} device, event, and scenario.
As \toolseqnature{} expects \symboltrafficsamples{} separate \trafficsample{}s, we split the full PCAP file using the event timestamps provided as part of the \datasetpingpong{} dataset.
For each split, we include 15 seconds of traffic following the event timestamps in the resulting \trafficsample{}.
This duration is chosen to stay consistent with~\cite{trimananda2020}.

The \datasetpingpong{} dataset spans 19 \iot{} devices, and there are between two and six different events per device.
Some devices are tested in different scenarios: local phone, remote phone, and/or IFTTT.
In the local (remote) phone scenario, the event is triggered using the companion app of the \iot{} device from a smartphone that is (not) part of the same LAN as the \iot{} device.
In the IFTTT scenario, the event is triggered using IFTTT~\cite{iftttwebsite}.
When extracting fingerprints from the \datasetpingpong{} dataset using \toolseqnature{}, we attempt to extract a fingerprint for each combination of \iot{} device, event, and scenario, \ie{} each such combination is analogous to the notion of an event $e$ used throughout Sections~\ref{sec:seqnature-fingerprinting-framework} and~\ref{sec:seqnature-fingerprinting-techniques}.
There are 140 such combinations.
While the \datasetpingpong{} dataset contains 50 invocations of each event, we only consider the first 10 invocations in order to keep the number of traffic samples consistent across datasets.
The \datasetpingpong{} dataset, as considered here, therefore encompasses 1,400 \trafficsample{}s.


\section{Fingerprinting Results}
\label{sec:seqnature-fingerprinting-results}
\toolseqnature{} facilitates studies of the relative performance of different fingerprinting techniques by (i) making it easy to implement a variety of fingerprinting techniques; and (ii) providing functionality that searches a dataset for manifestations of each fingerprint.
This section illustrates this idea by comparing the performance of the five fingerprinting techniques described in Section~\ref{sec:seqnature-fingerprinting-techniques}.
We first report the number of fingerprints discovered by each fingerprinting technique in Section~\ref{sec:seqnature-fingerprinting-results-prevalence}.
Then, in Section~\ref{sec:seqnature-fingerprinting-results-distinctiveness}, we examine if the extracted fingerprints are distinct among other traffic by analyzing how many false positives they give rise to in a closed-world scenario.

\begin{table*}[t!]
    \centering
    \begin{tabular}{|l|r|r|r|r|r|r|}
        \hline
        \multicolumn{1}{|c|}{\textbf{Dataset}} & \multicolumn{1}{c|}{\textbf{Events}} & \multicolumn{5}{c|}{\textbf{Prevalence}} \\
        & & \multicolumn{1}{c}{\sdbf{}s} & \multicolumn{1}{c}{\esdbf{}s} & \multicolumn{1}{c}{\ebf{}s} & \multicolumn{1}{c}{\fqdnbf{}s} & \multicolumn{1}{c|}{\esldbf{}s}\\
        \hline
        \datasetfingerprintv{}: \appletv{} & 1000 & 96\% \phantom{1}(964)& 96\% \phantom{1}(958) & 96\% \phantom{1}(959) & 95\% \phantom{1}(947) & 95\% \phantom{1}(947) \\
        \datasetfingerprintv{}: \firetv{} & 1000 & 100\% \phantom{1}(999) & 99\% \phantom{1}(993) & 100\% \phantom{1}(997) & 87\% \phantom{1}(865) & 87\% \phantom{1}(865) \\
        \datasetfingerprintv{}: \roku{} & 1000 & 100\% (1000) & 100\% (1000) & 100\% (1000) & 100\% (1000) & 100\% (1000) \\
        \hline
        \datasetpingpong{} & 140 & 68\% \phantom{11}(95) & 66\% \phantom{11}(92) & 81\% \phantom{1}(113) & 31\% \phantom{10}(44) & 32\% \phantom{10}(45) \\
        \hline
    \end{tabular}
    \caption{Prevalence of different fingerprint types in the \datasetfingerprintv{} and \datasetpingpong{} datasets. The prevalence is the percentage of events in the dataset that exhibit a fingerprint.}
    \label{tab:seqnature-prevalence}
\end{table*}

\subsection{Prevalence}
\label{sec:seqnature-fingerprinting-results-prevalence}
We start by reporting how many of the events in each dataset from Section~\ref{sec:seqnature-datasets} can be fingerprinted using each fingerprinting technique from Section~\ref{sec:seqnature-fingerprinting-techniques}.
For this, we adopt the nomenclature used in~\cite{varmarken2022fingerprintv}: the \emph{prevalence} is the percentage of events in the respective dataset that exhibit the respective type of fingerprint.
We also compare our results to those reported in~\cite{varmarken2022fingerprintv} and~\cite{trimananda2020} where applicable, \ie{} when a fingerprinting technique from Section~\ref{sec:seqnature-datasets} is a \toolseqnature{}-based approximation of a fingerprinting technique from these prior works.

Table~\ref{tab:seqnature-prevalence} reports the prevalence of \sdbf{}s, \esdbf{}s, \ebf{}s, \fqdnbf{}s, and \esldbf{}s in the datasets from Section~\ref{sec:seqnature-datasets}.
We discuss the results for each dataset below.

\parheader{\datasetfingerprintv{}.}
In general, there is little difference in the prevalence of the different types of fingerprints within and across \smarttv{} platforms: nearly all \smarttv{} apps exhibit all five types of fingerprints.
The exception is \firetv{}, where \fqdnbf{}s and \esldbf{}s are less prevalent (87\%).

The prevalence of \ebf{}s is 100\% for \firetv{}, which implies that a subset of \firetv{} apps (13\%) have \ebf{}s that exclusively consist of IP addresses.
This may be a result of developers hardcoding IP addresses, DNS caching across \trafficsamples{} in~\cite{varmarken2022fingerprintv}, or long-lived connections that span all \trafficsamples{} in~\cite{varmarken2022fingerprintv}.
The latter appears to be the primary reason as many of the IP addresses in these \ebf{}s belong to Amazon, which may in turn mean that some of these \ebf{}s actually capture OS-level activity rather than app-level activity.

We observe almost perfect agreement between the prevalence of \fqdnbf{}s and the prevalence of \dbfdefinition{}s (\dbf{}s) reported in~\cite{varmarken2022fingerprintv}.
The 1\% difference for \appletv{} and \firetv{} can be attributed to how the set of domains are constructed: domains in \fqdnbf{}s are (only) based on \tcp{} connections that are preceded by a corresponding \dns{} resolution or that carry a \tls{} \sni{} record, whereas \dbf{}s are constructed from \dns{} resolutions, irrespective of whether or not a \tcp{} connection is established to the domain, \tls{} \sni{}, and the host field of the \http{} header.

The astute reader may question why \esdbf{}s are more prevalent than \pbfdefinition{}s (\pbf{}s)~\cite{varmarken2022fingerprintv}, as \esdbf{}s require the endpoint, the packet sizes, and the packet directions to stay consistent across \trafficsamples{}, whereas \pbf{}s only require the packet sizes and the packet directions to stay consistent.
The explanation is that a \pbf{} only consists of the packet-pairs that appear \symboltrafficsamples{} times in total across the \symboltrafficsamples{} \trafficsamples{} (see Definition 4.2 in~\cite{varmarken2022fingerprintv}).
In contrast, an \esdbf{} only requires a packet sequence to occur in at least \symbolmintrafficsamples{} \trafficsamples{}, but does not constrain how many times the packet sequence is allowed to occur within each \trafficsample{} or in total across the \symboltrafficsamples{} \trafficsamples{}.

\parheader{\datasetpingpong{}.}
All five types of fingerprints are much less prevalent in the \datasetpingpong{} dataset: the highest prevalence we observe is 81\% (\ebf{}s), and the lowest is 31\% (\fqdnbf{}s).
However, some devices in the \datasetpingpong{} dataset do not generate external \tcp{} traffic for some event and scenario combinations (recall from Section~\ref{sec:seqnature-fingerprinting-technique-preprocessing} that our focus in this work is on fingerprints that are observable upstream from the home router's WAN port).
Naturally, these events are impossible to fingerprint using any fingerprinting technique that operates on external \tcp{} traffic.
If we only consider events where the target device exchanges at least one \tcp{} packet with an Internet endpoint in every \trafficsample{}, then the total number of events in the \datasetpingpong{} dataset decreases from 140 to 116.
With this baseline, the prevalence becomes 82\%, 79\%, 97\%, 38\%, and 39\% for \sdbf{}s, \esdbf{}s, \ebf{}s, \fqdnbf{}s, and \esldbf{}s, respectively.

Since an \sdbf{} is a \toolseqnature{}-based approximation of a \plsdefinition{} (\pls{}), we compare the number of \sdbf{}s extracted to the number of \pls{}s identified for events with external traffic~\cite{trimananda2020}.
\toolseqnature{} identifies \sdbf{}s for 95 events, while the \toolpingpong{} tool only identifies \pls{}s for 78 events. 
\sdbf{}s are expected to be more prevalent than \pls{}s as \pls{}s are more restrictive: packet-pairs are discarded if they occur multiple times, and packet sequences must occur in the same order across \trafficsamples{}.

\subsection{Distinctiveness}
\label{sec:seqnature-fingerprinting-results-distinctiveness}
We now compare the discriminative power of the different types of fingerprints by examining how distinct the extracted fingerprints are from other traffic.

We refer to a manifestation of the fingerprint, \fofapp{e_1}, of event $e_1$ in a \trafficsample{} for event $e_2$, where $e_1 \ne e_2$, as a false positive.
We compare the discriminative power of the fingerprints in Table~\ref{tab:seqnature-prevalence} by (i) searching all \trafficsamples{} of each dataset for manifestations of each and every fingerprint; and then (ii) summarizing how many false positives the different types of fingerprints give rise to.
This false positive analysis is a more comprehensive evaluation of the fingerprints' discriminative power than prior work~\cite{varmarken2022fingerprintv}, 
where each app's fingerprint is only compared to other apps' fingerprints.
In contrast, we compare each fingerprint against \emph{all} traffic in each dataset, including traffic that does not find its way into any fingerprint.

In our analysis below, we report the number of distinct events the false positives are spread across.
Recall from Section~\ref{sec:seqnature-datasets} that each dataset contains $\symboltrafficsamples{}=10$ \trafficsamples{} for each event.
Now, assume that we identify 10 false positives for the fingerprint, \fofapp{e}, of some event $e$.
These 10 false positives could stem from manifestations of \fofapp{e} in 10 \trafficsamples{} from 10 different events, or from \fofapp{e} consistently appearing in all \trafficsamples{} of a single event, or somewhere in between these two extremes.
Thus, if we only consider the total number of false positives per fingerprint, it becomes difficult to discern if the false positives are spread across \trafficsamples{} from many different events, or if they repeatedly occur in most/all \trafficsamples{} of a single/few event(s).
In our view, the number of different events the false positives are spread across is therefore a better measure of how distinct a fingerprint is among a wide variety of other traffic.

\begin{figure*}
    \centering
    \begin{subfigure}{\columnwidth}
        \includegraphics[width=\columnwidth]{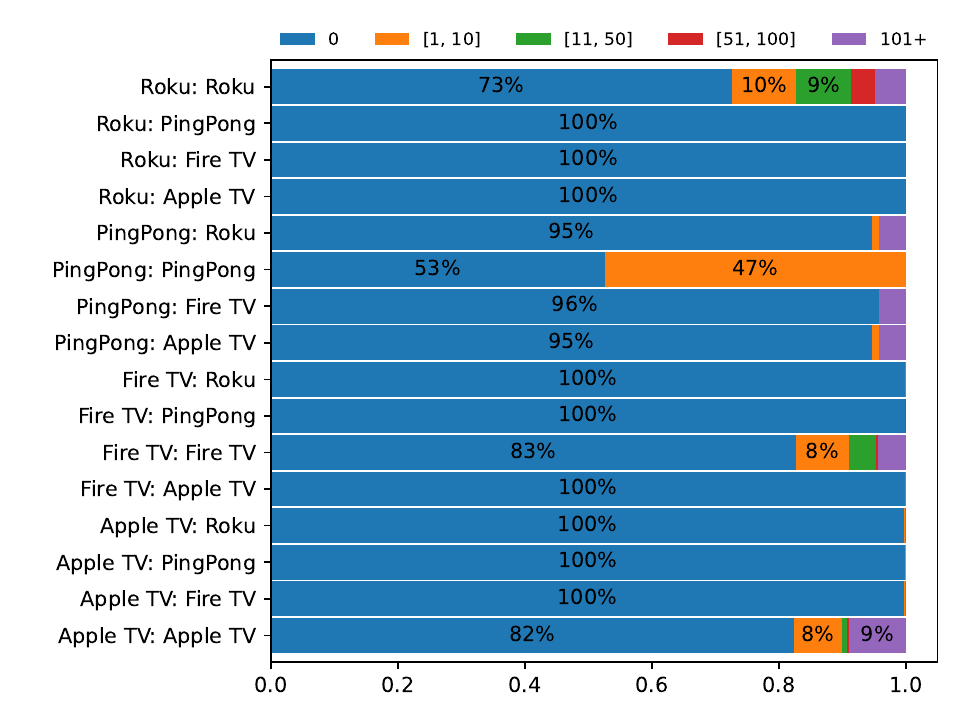}%
        \caption{\sdbfdefinition{}s (\sdbf{}s)}%
        \label{fig:seqnature-false-positives-sdbf}%
    \end{subfigure}\hfill%
    \begin{subfigure}{\columnwidth}
        \includegraphics[width=\columnwidth]{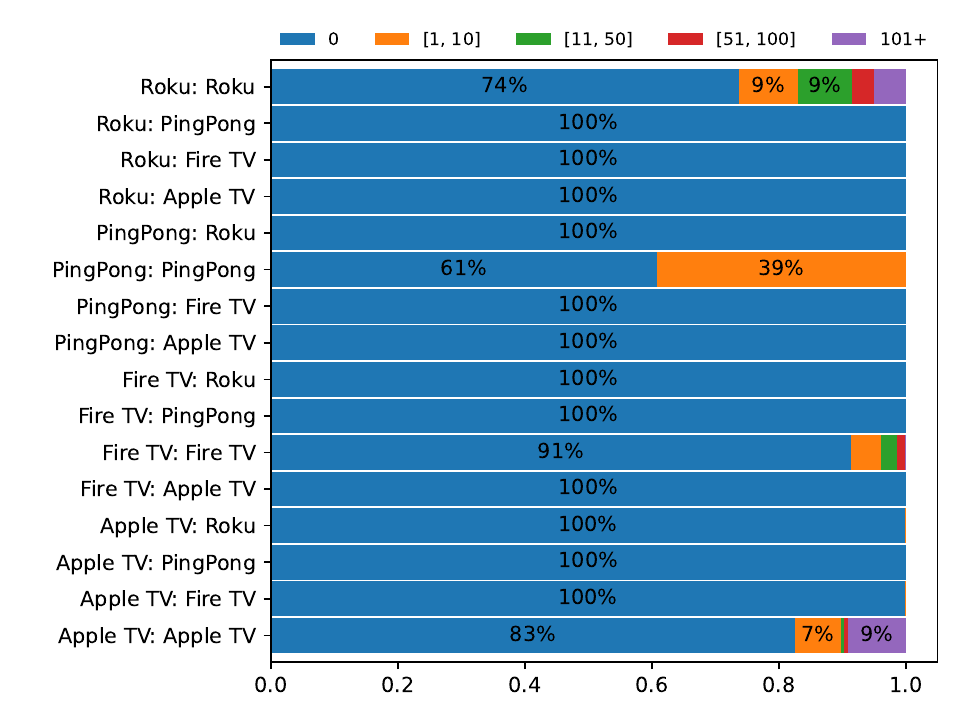}%
        \caption{\esdbfdefinition{}s (\esdbf{}s)}%
        \label{fig:seqnature-false-positives-esdbf}%
    \end{subfigure}\hfill%
    \begin{subfigure}{\columnwidth}
        \includegraphics[width=\columnwidth]{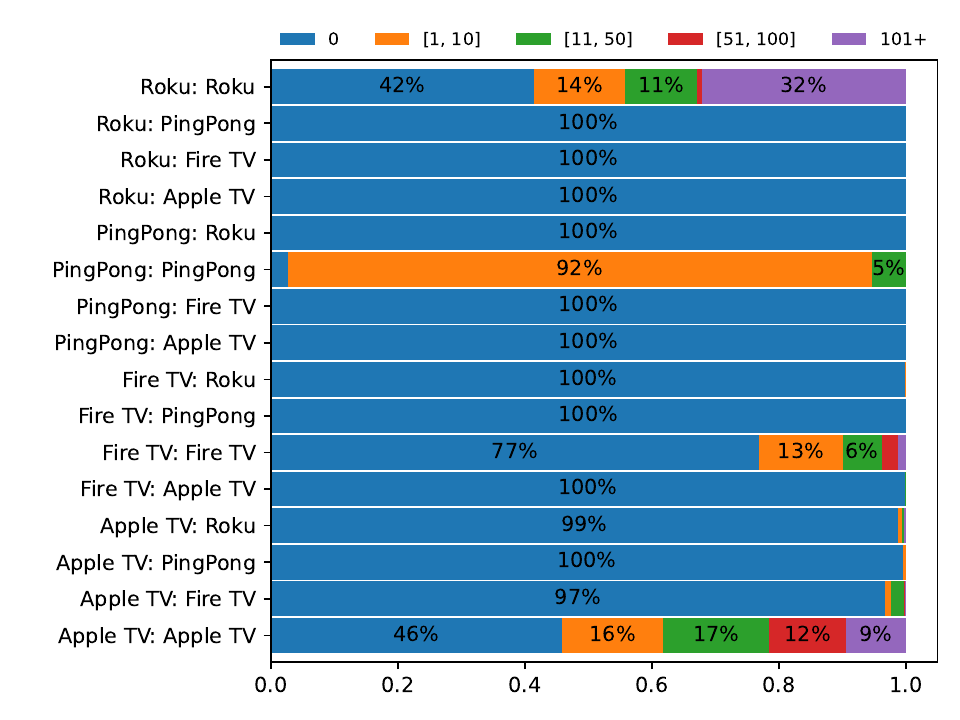}%
        \caption{\ebfdefinition{}s (\ebf{}s)}%
        \label{fig:seqnature-false-positives-ebf}%
    \end{subfigure}\hfill%
    \begin{subfigure}{\columnwidth}
        \includegraphics[width=\columnwidth]{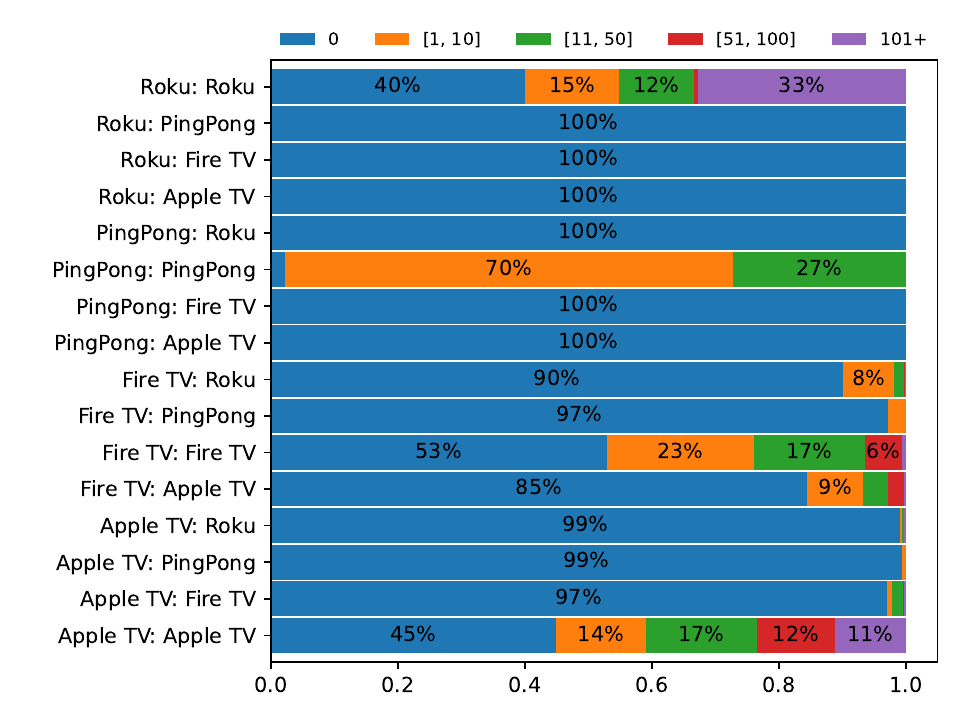}%
        \caption{\fqdnbfdefinition{}s (\fqdnbf{}s)}%
        \label{fig:seqnature-false-positives-fqdnbf}%
    \end{subfigure}\hfill%
    \begin{subfigure}{\columnwidth}
        \includegraphics[width=\columnwidth]{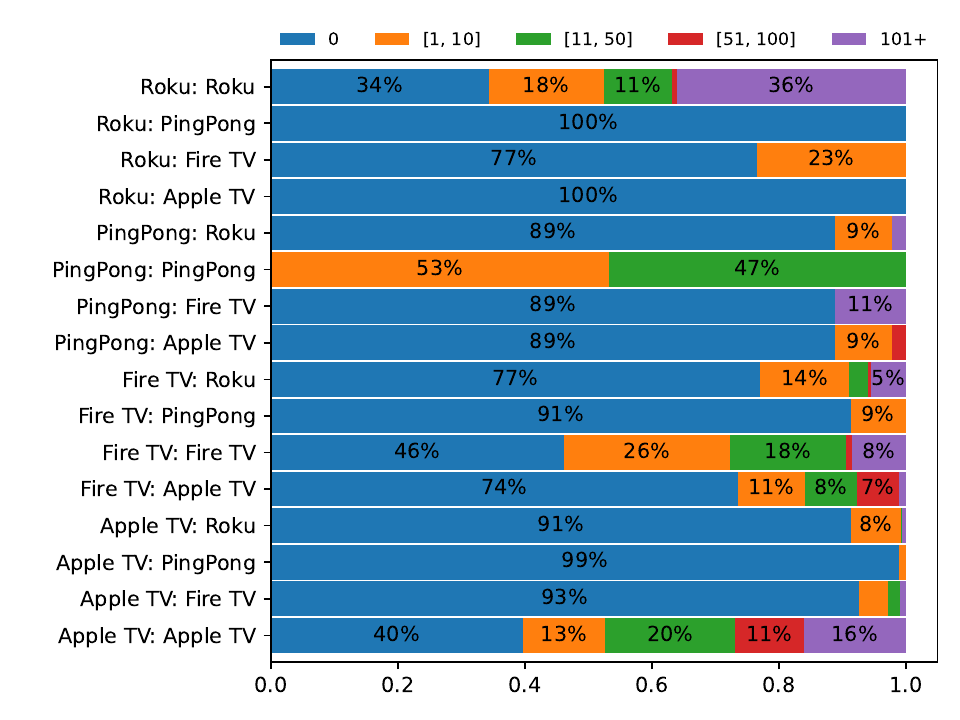}%
        \caption{\esldbfdefinition{}s (\esldbf{}s)}%
        \label{fig:seqnature-false-positives-esldbf}%
    \end{subfigure}%
    \caption{Breakdown of how many fingerprints give rise to false positives in how many events. The notation $d_1\mathrm{:} d_2$ on vertical axes denotes false positives of fingerprints extracted from dataset $d_1$ observed in dataset $d_2$. Example (Figure~\ref{fig:seqnature-false-positives-sdbf}): when \sdbf{}s extracted from the \datasetpingpong{} dataset are evaluated against the \datasetpingpong{} dataset (``\datasetpingpong{}: \datasetpingpong{}''), 53\% of \sdbf{}s produce 0 false positives, while 47\% of \sdbf{}s produce false positives in traffic that stems from between 1 and 10 different events.}
    \label{fig:seqnature-false-positives}
\end{figure*}

Figure~\ref{fig:seqnature-false-positives} provides a breakdown of how many fingerprints give rise to false positives in how many events (the number of events have been binned to make the plots legible).
Across all fingerprint types, we see a tendency that the majority of false positives occur within-platform, \eg{} fingerprints for \appletv{} apps evaluated against traffic from other \appletv{} apps.
We surmise that this may in part be due to apps' use of platform-specific cloud infrastructure.

\parheader{\sdbf{}s and \esdbf{}s.}
Across all datasets, the majority of \sdbf{}s and \esdbf{}s produce no false positives  (the 0 bin makes up at least 53\% of every bar in Figures~\ref{fig:seqnature-false-positives-sdbf} and~\ref{fig:seqnature-false-positives-esdbf}).

The \sdbf{}s that were extracted from the \datasetpingpong{} dataset appear to have least discriminative power, as 47\% of these \sdbf{}s produce at least one false positive when evaluated against the \datasetpingpong{} dataset, and a small fraction (<5\%) also produce 101+ false positives in the \appletv{}, \firetv{}, and \roku{} subsets of the \datasetfingerprintv{} dataset.
At face value, this result may seem surprising, as \pls{}s extracted from the \datasetpingpong{} dataset produced very few false positives~\cite{trimananda2020}.
However, as discussed in Section~\ref{sec:seqnature-fingerprinting-results-prevalence}, \pls{}s set stricter requirements than \sdbf{}s w.r.t. what packet sequences are selected for inclusion in the fingerprint and how these packet sequences are ordered.
As shown in Section~\ref{sec:seqnature-fingerprinting-results-prevalence}, this resulted in the discovery of more \sdbf{}s than \pls{}s, but---given the false positives results discussed above---it appears that at least some of these additional \sdbf{}s are rather weak fingerprints.

By comparing \sdbf{}s to \esdbf{}s (Figures~\ref{fig:seqnature-false-positives-sdbf} and~\ref{fig:seqnature-false-positives-esdbf}), we observe that the introduction of the endpoint as an additional feature results in a slight decrease in overall false positives.
The effect is most pronounced for fingerprints extracted from the \datasetpingpong{} dataset, where it helps eliminate false positives in the \appletv{}, \firetv{}, and \roku{} subsets of the \datasetfingerprintv{} dataset.

\parheader{\ebf{}s, \fqdnbf{}s, and \esldbf{}s.}
A large fraction of \ebf{}s produce false positives when evaluated against traffic from the same platform.
For example, 97\% of \ebf{}s extracted from the \datasetpingpong{} dataset produce at least one false positive when evaluated against the \datasetpingpong{} dataset.
This result is in agreement with prior work that has shown that while \iot{} devices can be identified from the endpoints they contact, other information, such as traffic rates or individual packet exchanges, is necessary to differentiate between different events on the same device~\cite{apthorpe2017, oconnor2019homesnitch, trimananda2020, acar2020peekaboo}. 

When comparing \ebf{}s to \fqdnbf{}s (Figures~\ref{fig:seqnature-false-positives-ebf} and~\ref{fig:seqnature-false-positives-fqdnbf}), we observe that the elimination of IP addresses from the fingerprints results in a slight increase in false positives.
The effect is most noticeable for \firetv{}: the fraction of fingerprints that produce any number of false positives doubles within-platform, and false positives also start to appear in the \appletv{} and \roku{} subsets of the \datasetfingerprintv{} dataset, and in the \datasetpingpong{} dataset.

From comparing \fqdnbf{}s to \esldbf{}s (Figures~\ref{fig:seqnature-false-positives-fqdnbf} and~\ref{fig:seqnature-false-positives-esldbf}), it becomes clear that the granularity used for the domain name has a significant effect on how distinct fingerprints are from other traffic: unlike \fqdnbf{}s, where cross-dataset false positives were mostly limited to \firetv{} only, false positives occur across the board for \esldbf{}s, except for when \roku{} \esldbf{}s are evaluated against the \appletv{} subset of the \datasetfingerprintv{} dataset and the \datasetpingpong{} dataset.

We also compare our false positive results for \fqdnbf{}s to the distinctiveness of \dbf{}s reported in~\cite{varmarken2022fingerprintv}.
When the \fqdnbf{}s for \appletv{}, \firetv{}, and \roku{} are evaluated within-platform, only 45\%, 53\%, and 40\% of \appletv{}, \firetv{}, and \roku{} \fqdnbf{}s, respectively, produce no false positives.
In contrast, 59\%, 63\%, and 46\% of \appletv{}, \firetv{}, and \roku{} \dbf{}s, respectively, were found to be distinct~\cite{varmarken2022fingerprintv}.
The difference illustrates that while the methodology for evaluating fingerprint distinctiveness proposed in~\cite{varmarken2022fingerprintv} is efficient, it can only establish an optimistic upper bound on distinctiveness as it only compares fingerprints to other fingerprints, but not to non-fingerprint traffic. 
Our false positive analysis provides a more accurate estimate of fingerprints' distinctiveness, but at a higher computational cost.

\section{Limitations and Future Directions}
\label{sec:seqnature-limitations-and-future-directions}
This section discusses the limitations of our work and outlines future directions.
In summary, we plan to analyze false positives in further detail, and to use \toolseqnature{} to implement and evaluate additional fingerprinting techniques.

\parheader{False Positives in \nat{}'ed Traffic from Multiple Devices.}
A limitation of the closed-world false positive tests in Section~\ref{sec:seqnature-fingerprinting-results-distinctiveness} is that each fingerprint is only compared against traffic from a single event at a time.
Most gateways in real-world home networks use \natdefinition{} (\nat{}), which makes traffic from multiple devices appear as traffic from a single device.
This increases the potential for false positives, as a false positive will arise whenever the union of traffic from all devices in the home network contains a manifestation of the fingerprint, even when no single device generates traffic that matches the fingerprint.

We leave a more exhaustive evaluation of false positives to future work as our objective here is not to definitively rank the different types of fingerprints \wrt{} their respective costs and benefits, but rather to demonstrate how \toolseqnature{} facilitates comparisons of different fingerprinting techniques.
To simulate NAT'ed traffic, one option that future work may pursue is to combine \trafficsamples{} from a set of events $E$ into a single \trafficsample{} and then examine the resulting \trafficsample{} for false positives of the fingerprint, \fofapp{e}, of event $e$, where $e \notin E$.

\parheader{Fingerprints with Many False Positives.}
Across all five fingerprint types, a small fraction of fingerprints produce many false positives, as evident from the 101+ bins in Figure~\ref{fig:seqnature-false-positives}.
We plan to examine these fingerprints in more detail to understand why they have little discriminative power.

Preliminary observations for \esdbf{}s indicate that if an \esdbf{} of a \smarttv{} app gives rise to many false positives, it is generally because the \esdbf{} only comprises a few packet sequences that are exchanged with platform-specific endpoints.
These packet exchanges likely stem from the app's use of operating system (OS) services that incur network activity.
If an \esdbf{} only consists of these packet exchanges, it will manifest itself in traffic from all other apps that also utilize these OS services.
Notice that this also helps explain why most false positives for \esdbf{}s are within-platform: if most apps consistently utilize these OS services, then the resulting packet exchanges will become part of most apps' \esdbf{}s, which will make the \esdbf{}s distinct from traffic of apps on other platforms, as apps on other platforms will not utilize these platform-specific OS services.

\parheader{Exploring Fingerprinting Techniques.}
In this paper, we have examined five types of fingerprints that broadly fall into two categories: (i) fingerprints that consider the amount of data being exchanged in each direction, at the granularity of individual packets (\sdbf{}s and \esdbf{}s); and (ii) fingerprints that only consider what servers are contacted and pay no attention to what data is exchanged with said servers (\ebf{}s, \fqdnbf{}s, and \esldbf{}s).

Going forward, we intend to use \toolseqnature{} to explore additional new fingerprinting techniques that fall between these two extremes.
This could for example include fingerprints that consider the total traffic volume (in each direction) across the first \symbolstreamprefixlength{} packets of each stream, or fingerprints that consider the total number of upstream/downstream packets across the first \symbolstreamprefixlength{} packets of each stream.
\toolseqnature{}'s configurability, in particular the option to provide a user-defined distance metric, makes it simple to implement such fingerprinting techniques. 
For example, to implement a fingerprinting technique that only considers the total traffic volume, one merely needs to implement a distance metric that computes the distance between two packet sequences $p_1$ and $p_2$ as $\left| \left( \sum_{i=1}^{n}{\left|p_1[i]\right|} \right) - \left( \sum_{i=1}^{n}{\left|p_2[i]\right|} \right) \right|$, where $n$ is the length of $p_1$ and $p_2$, and $\left|p_x[i]\right|$ is the size of the $i$'th packet in packet sequence $p_x$.

We envision that \toolseqnature{} will provide a universal framework for specifying and evaluating existing, and discovering novel fingerprints. \toolseqnature{} can be applied to different applications and datasets and enable its user to discover, customize, and fine-tune fingerprints to a particular application context. For example, packet-pair sizes have been shown to be a universal fingerprint for simple \iot{} devices~\cite{trimananda2020}; the sizes of the two packets in a pair vary per device but can be automatically discovered and extracted by \toolseqnature{}. Conversely, the same methodology can be used to inform the design of network protocols that makes it difficult for an adversary to fingerprint end devices and their events.

\section{Conclusion}
\label{sec:seqnature-summary}
In this paper, we presented \toolseqnature{}, a general fingerprinting framework that makes it simple to implement and evaluate, in a unifying way, any arbitrary fingerprinting technique that extracts network fingerprints of events, based on packet sequences.
We demonstrated the versatility of \toolseqnature{} by implementing five different fingerprinting techniques, including two that approximate well-known fingerprinting techniques from prior work.
We applied the five fingerprinting techniques to network traffic datasets from the literature, extracting fingerprints for \smarttv{} apps and events on simple \iot{} devices.
We examined the discriminative power of the extracted fingerprints by relying on \toolseqnature{}'s ability to search network traffic for false positives.
By comparing the number of false positives across fingerprint types, we offered insights into the importance of (the granularity of) different features.
Our findings also confirmed some results from prior work, including that endpoint information alone is insufficient to distinguish between individual events on \iot{} devices, but also showed that \smarttv{} app fingerprints based exclusively on endpoint information are not as distinct as previously reported.

Future work can take similar steps to compare the performance of different fingerprinting techniques. We envision that \toolseqnature{} will provide a universal framework for specifying and evaluating existing, and discovering novel fingerprints. \toolseqnature{} can be applied to different applications and datasets and enable its user to discover, customize, and fine-tune fingerprints to a particular application context. Conversely, the same methodology can be used to inform the design of network protocols so as to make it more difficult for an adversary to fingerprint end devices and their events. To this end, we plan to make \toolseqnature{} publicly available.

\section*{Acknowledgments}
This work was partially supported by NSF Awards 19563931 and 1900654.
The authors would like to thank Maganth Seetharaman and Amogh Pradeep for assisting with explorative data collection, and Andrew Searles for discussions on data analysis.



\bibliographystyle{plain}
\bibliography{references.bib}

\end{document}
